\documentstyle[preprint,fleqn,aps,prc]{revtex}
\newcommand{\cfig}[3]{%
	\begin{figure}[tbp]%
        \caption[]{\label{#1} #2}%
	\end{figure}
}
\newcommand{\iunit}{{\rm i}}

\newcommand{\fm}{\mbox{fm}}
\newcommand{\barn}{\mbox{b}}
\newcommand{\mev}{\mbox{MeV}}
\newcommand{\plv}{\mbox{PL--40}}
\newcommand{\nls}{\mbox{NL--SH}}
\newcommand{\nuc}{{\rm n}}

\newcommand{\ie}{{\em i.~e.}}
\newcommand{\eg}{{\em e.~g.}}

\newcommand{\empcol}{\multicolumn{2}{c}{}}
\newcommand{\lag}{{\cal L}}
\newcommand{\nn}{\nonumber}
\newcommand{\ffrac}[2]{{\textstyle\frac{#1}{#2}}}
\newcommand{\half}{\ffrac{1}{2}}

\newcommand{\Psibar}{\overline{\Psi}}

\newcommand{\scalar}[2]{#1\!\cdot\!#2}
\newcommand{\plut}{{${}^{240}$Pu}}
\newcommand{\radi}{{${}^{226}$Ra}}
\newcommand{\thor}{{${}^{232}$Th}}

\newcommand{\be}{\begin{equation}}
\newcommand{\ee}{\end{equation}}
\newcommand{\ben}{\begin{displaymath}}
\newcommand{\een}{\end{displaymath}}
\newcommand{\bea}{\begin{eqnarray}}
\newcommand{\eea}{\end{eqnarray}}

\newcommand{\reffig}[1]{Figure~\ref{#1}}

\begin{document}
\draft
\title{Fission barriers and asymmetric ground states
in the relativistic mean field theory}
\author{K. Rutz${}^{a}$, J. A. Maruhn${}^{a,c}$,
        P.--G. Reinhard${}^{b,c}$, and W. Greiner${}^{a,c}$}
\address{${}^{a}$ Institut f\"ur Theoretische Physik, 
         Universit\"at Frankfurt\\
         Robert-Mayer-Str. 10, D-60325 Frankfurt, Germany.}
\address{${}^{b}$ Institut f\"ur Theoretische Physik, 
         Universit\"at Erlangen\\
         Staudtstr.\ 7, D-91058 Erlangen, Germany}
\address{${}^{c}$ Joint Institute for Heavy-Ion Research, 
         Holifield Heavy Ion Research Facility\\
         Oak Ridge, TN 37831, U.S.A.}
\date{5 January 1995}
\maketitle
\begin{abstract}
The symmetric and asymmetric fission path for \plut, \thor, and
\radi\
is investigated within the relativistic mean--field model. Standard
parametrizations which are well fitted to nuclear ground state
properties are found to deliver reasonable qualitative and
quantitative features of fission, comparable to similar nonrelativstic
calculations. Furthermore, stable octupole deformations
in the ground states of Radium isotopes are investigated. They are
found in a series of isotopes, qualitatively in agreement with
nonrelativistic models. But the quantitative details differ amongst
the models and between the various relativsitic parametrizations.
\end{abstract}
        
\section{Introduction}
Nuclear fission has been the challenge to develop microscopic theories of
nuclear collective motion. Only a subtle interplay of collective
deformations and changing microscopic shell structure can explain the
quantitative details like tunneling times and fragment mass distributions
\cite{FunHills}. One of the most decisive features to pin down
the contributions from the shell effects is the occurence of 
multiple--humped barriers \cite{BjoDHB}. The macroscopic--microscopic method,
based on a phenomenologically fitted shell model plus Strutinsky
shell corrections \cite{Strut}, allowed quite early extensive investigations
of the various landscapes of collective Potential Energy Surfaces
(PES), see \eg\ \cite{FunHills,Mol72,Web76,Mol81,Lea82}.
Since the early seventies, fully selfconsistent mean--field
models have been available for nuclei. These are the nonrelativistic
Hartree--Fock models with the Skyrme force \cite{Skyrme,Vau72}
or the Gogny force \cite{Gog74}. At about the same time competitive
relativistic mean--field models for nuclear structure were proposed
\cite{Wal74,BogBod}. Although much more elaborate, selfconsistent
mean--field calculations of the PES for nuclear fission
appeared rather soon for the Skyrme--Hartree--Fock
models \cite{FloFis} and later for the Gogny force \cite{GogFis},
it was only recently that PES for fission have been
calculated with the relativistic mean--field model \cite{Blu94}.
These calculations have shown that well adjusted parametrizations
for the relativistic mean--field model, see \eg\ \cite{ReiRuf,Rei89},
deliver reasonable fission barriers comparable to those of
nonrelativistic calculations. A variation of the parametrization,
in particular with respect to the effective mass, has shown that the
requirement to reproduce reasonable fission barriers excludes many
variants and only the standard best--fit forces with the typically very
low effective nucleon mass remain. It is the aim of this paper to
extend the investigations of \cite{Blu94} to asymmetric fission and
to study a larger variety of actinides. We confine the consderations
to the two parametrizations NL1 from \cite{ReiRuf} and \plv\ from
\cite{Rei88} which have proven to provide reasonable fission barriers
\cite{Blu94}. In addition, we consider the parametrization \nls\
which is claimed to be better adjusted with respect to isovector
properties \cite{Sha93}. As a second objective of this paper,
we exploit the asymmetric degree of freedom in our calculations to
investigate the possibility of stable octupole deformations in the
ground states of Radium isotopes which, again, have been investigated
first in the macroscpic--microscopic method \cite{Mol81,Lea82}.

The paper is outlined as follows: In section~\ref{sectheo}, we give
a short account of the model, its parameters and further ingredients.
In section~\ref{secfiss}, we present results for the PES for symmetric
and asymmetric fission of \plut, \thor, and \radi. And in
section~\ref{secoctu}, we discuss the stable octupole deformations
in the ground states of Radium isotopes.

\section{Theoretical background}
\label{sectheo}

The relativistic mean--field model is meanwhile a standard
in nuclear physics, for detailed reviews see \cite{SerWal,ShakRev,Rei89}.
Thus there is not much to explain. But for completeness we specify
here the Lagrangian of the model. It reads
\bea
\lag_{\rm RMF}&=&
\lag_{\rm nucleon}^{\rm free}+\lag_{\rm meson}^{\rm free}+
\lag_{\rm coupl}^{\rm lin}+\lag_{\rm coupl}^{\rm nonlin}\nn\\
\lag_{\rm nucleon}^{\rm free}&=&
\Psibar\left(\iunit\gamma_\mu\partial^\mu-m_\nuc\right)\Psi\nn\\
\lag_{\rm meson}^{\rm free}&=&\half\left(\partial_\mu\Phi\partial^\mu\Phi-
m_\sigma^2\Phi^2\right)
-\half\left(\half G_{\mu\nu}G^{\mu\nu}-m_\omega^2 V_\mu
V^\mu\right)\nn\\
&&{}-\half\big(\half\scalar{\vec{B}_{\mu\nu}}{\vec{B}^{\mu\nu}}-
m_\rho^2\scalar{\vec{R}_\mu}{\vec{R}^\mu}\big)
-\ffrac{1}{4}F_{\mu\nu}F^{\mu\nu}\nn\\
\lag_{\rm coupl}^{\rm lin}&=&{}-g_\sigma\Phi\Psibar\Psi
-g_\omega V_\mu\Psibar\gamma^\mu\Psi
-g_\rho\scalar{\vec{R}_\mu}{\Psibar\vec{\tau}\gamma^\mu\Psi}
-e A_\mu\Psibar\ffrac{1+\tau_0}{2}\gamma^\mu\Psi\nn\\
\lag_{\rm coupl}^{\rm nonlin}&=&\half m_\sigma^2\Phi^2-U(\Phi)\nn
\eea
where $\Psi$ is the nucleon field, $\Phi$ the scalar-isoscalar field,
$V_\mu$ the vector-isoscalar field,
$\vec{R}_\mu$ the vector-isovector field, and
$A_\mu$ the photon field. The corresponding force tensors are
\ben
G_{\mu\nu}=\partial_\mu V_\nu -\partial_\nu V_\mu\ ,\quad
\vec{B}_{\mu\nu}=\partial_\mu \vec{R}_\nu -\partial_\nu\vec{R}_\mu\ ,\quad
F_{\mu\nu}=\partial_\mu A_\nu -\partial_\nu A_\mu\ .
\een
The $U(\Phi)$ is the nonlinear functional for the scalar field.
We consider two variants, first the standard nonlinear functional
\be
\label{Ustandard}
  U(\Phi)=
  \half m_\sigma^2\Phi^2+\ffrac{1}{3}b_2\Phi^3+\ffrac{1}{4}b_3\Phi^4
  \ ,
\ee
and second, the stabilized nonlinear functional \cite{Rei88}
\bea
\label{Ustab}
  U(\Phi) &=& \ffrac{1}{2} m_\infty^2\Phi^2\nn\\
  &&+\Delta m\Big\{ \ffrac{\delta\Phi^2}{2}
     \left[
       \log\left(1+\left(\ffrac{\Phi-\Phi_0}{\delta\Phi}\right)^2\right)
       -\log\left(1+\left(\ffrac{\Phi_0}{\delta\Phi}\right)^2\right)
     \right] \nn \\
  &&\qquad
   +\Phi_0\Phi\left(1+\left(\ffrac{\Phi_0}{\delta\Phi}\right)^2\right)^{-1}
    \Big\} \ .
\eea
The stabilized functional is necessary whenever the shell fluctuations
are likely to produce large peaks in the the scalar density. This
happens in light nuclei \cite{Rei88} and at large deformations
\cite{BluFin}. There is a large variety of parametrizations around
in the literature, see \cite{Rei89}.
We will discuss here three different sets: NL1 and \nls\ within
the standard nonlinear model (\ref{Ustandard}) and \plv\ within
the stabilized variant (\ref{Ustab}). The parameters are listed
in Table~\ref{NLsets} and \ref{stabsets}.
The sets NL1 and \plv\ are obtained from a fit to ground state properties
of spherical nuclei, as explained in \cite{ReiRuf} for NL1 and
\cite{Rei88} for \plv. These fits
take care of the nuclear charge form factor in terms of the diffraction
radius and a surface thickness. The set \nls\ has been obtained also
from a fit, but biased more on a smaller isovector parameter and
employing only the r.m.s.~radii as global information in the nuclear shape.
It is claimed that this set is more appropriate for exotic nuclei
due its smaller, and thus more realistic, isovector strength
\cite{Sha93}. It is to be remarked, however, that the fit of \nls\
includes less information on the nuclear shape because only the overall
extension in terms of the r.m.s. radius was employed whereas NL1 and
\plv\ carry some extra information on the surface thickness and 
thus have probably more reliable surface properties.
The parameters of the model, as given in Table~\ref{NLsets}
and \ref{stabsets}, serve to specify the used models in detail.
More insight into the physical
properties of the model is provided by listing the parameters of
symmetric nuclear matter. This is done in Table~\ref{nucmat}
where we also provide the results for the comparable nonrelativistic
models and the experimental data. The most reliable data are the
binding energy and equilibrium density, which are well reproduced by all
sets. The incompressibility is less well known and accordingly there is
somewhat more variation amongst the forces. Where the effective mass is
concerned, it is to be noted that the definitions differ. The
relativistic models have a much lower $m^*/m$ throughout. But the
relevant quantity for nuclear structure calculations are the single
particle level densities at the Fermi surface, and these turn out to be
comparable amongst nonrelativistic and relativistic sets \cite{NazRei}.
The differences in the symmetry energy cannot be explained away. The
relativistic models have a tendency to overestimate it. The force \nls\
has a more realistic value because particular attention was paid to this
observable during the fit. It is to be noted that the symmetry energy
determines mainly the position of the isovector giant resonances. It is
yet an open question how it affects the isotopic trends in the nuclear
ground states.

All three parametrizations are designed to determine an appropriate
nuclear mean--field. They need to be completed by a recipe to define the
occupation of states. We have used pairing in the constant gap approach
with \cite{BloFlo}
\ben
  \Delta
  =
  11.2\,\mev/\sqrt{A}
  \ .
\een
This is, admittedly, a rather rough estimate.
Varying pairing recipes can change the fission barriers by about $\pm
1\,\mev$ \cite{BluDiss}. There are uncertainties in a similar order of
magnitude in other parts of the treatment, see below. We can thus live
with that level of approach for pairing in the present stage.

Finally, we take into account a correction for the spurious
centre--of--mass motion. The standard nonlinear sets NL1 and \nls\ used
the estimate
\be
  E_{\rm cm}
  =
  \ffrac{3}{4}\cdot 41\,A^{-1/3}\,\mev
\label{Ecmguess}
\ee
whereas \plv\ used a microscopically calculated
$
  E_{\rm cm}
  =
  \langle \hat{P}_{\rm cm}^2\rangle/(2Am_{\rm n})
  \ .
$
The microscopic evaluation is a bit tedious in
non--spherical codes. The estimate (\ref{Ecmguess}) is a fair
replacement, particularly in heavy nuclei as studied here.
Admittedly, the centre--of--mass correction alone is somewhat incomplete.
It suffices only for spherical nuclei. Deformed nuclei would require a
rotational projection as well, and collective dynamics like fission,
is only complete if also the (collective) vibrational zero--point
energies are carefully accounted for. This is a very demanding task.
In particular, it requires access to the appropriate collective cranking
masses which are not yet available. And a proper impementation of
cranking in the relativistic framework will be a much more demanding
task than in nonrelativstic mean--field models because the full Fermi sea
of occupied antinucleon states needs to be kept projected out. This
inhibits, \eg, the efficient linear response techniques on the grid
\cite{LRgrid,LRwf}. We therefore dismiss these details for the moment.
An estimate for all the effects from collective zero--point energies can
be taken from the two--centre shell model \cite{Rei78}: the first barrier
is lowered by 0.5\,\mev\ and the second barrier by 2\,\mev. This is thus the
uncertainty in our present calculations. It is about the same
order of magnitude as the uncertainty from other parts of the model, \eg\
the simple pairing recipe. In that sense, the present level of approach
is well equilibrated with respect to precision. However, when
comparing with other models, we will have to examine which zero--point
energies had been included there and counter--correct properly.

Besides the energy, we will consider several multipole moments
\ben
  \hat{Q}_\ell =
  \sum_i r_i^\ell Y_{\ell 0}(\Omega_i)
\een
as observables characterizing the shapes of the nuclei
and the electrical dipole moment
\ben
  \hat{D} = e\frac{NZ}{A}
  \Big(
  \frac{1}{Z}\sum_{i\in {\rm prot.}} z_i
  -\frac{1}{N}\sum_{i\in {\rm neut.}} z_i
  \Big)\ .
\een
For historical reasons we also display the cartesian quadrupole
moment
\ben
  \hat{Q} =
  \sqrt{\ffrac{16\pi}{5}} \hat{Q_2}\ .
\een
There are, in principle, two different sets of multipole moments
to be considered in a relativistic model, one computed with the
scalar density and another one computed with the vector density
(the electrical dipole, of course, is uniquely related to
the vector--isovector density). The differences are very small
in the present relativistic mean--field model
\cite{FinDiss}.
Thus any choice is possible. We are using here the multipole moments
from the isoscalar--vector density. Another observable deduced from the
isoscalar--vector density is the mass of the heavy fragment
\be
  A_{\rm h} =
  \int\limits_{z<z_{\rm neck}} {\rm d}V\,\rho_0(\vec{r})
  \qquad\mbox{for}\qquad
  \langle \hat{Q}_3\rangle > 0\ .
\label{massfrag}
\ee
It can be computed only for large deformations when the 
fragments start to develop visibly.

The multipole moments  serve also as
constraints in order to map the whole potential energy surfaces 
for deformation and fission. They are related to the vector
density and thus can be added as a potential--like term in
the effective Hamiltonian of the model, \ie
\ben
  \hat H^{\rm C} =
  \hat H - \sum_{\ell=1,3} \lambda_\ell \hat{Q}_\ell^{\rm D}\ ,
\een
were $\ell=2,3$ serves to implement a constraint on $\langle Q_2
\rangle$ and $\langle Q_3 \rangle$, and $\ell = 1$ is included to fix the
centre--of--mass at $\langle z_{\rm cm} \rangle=0$.
The upper index "D" denotes {\em damped} multipole moments.
Some damping is required because the mere multipole moments
increase rapidly towards the edges
of the numerical box
which causes several unpleasant numerical instabilities.
For example, every nucleus becomes asymptotically unstable
with the slightest quadrupole constraint because there arises
always one direction where the potential decreases as $-r^2$
with $r \rightarrow \infty$. The problem becomes worse with
the octupole constraint.
The multipole operators thus
have to be cut off at large distances from the nucleus. We do
this by multiplying them by a Fermi function
\ben
\label{dampC}
  \hat C_\ell^{\rm D} =
  \hat{C}_\ell
  \left[1+\exp\frac{\Delta R-\alpha}{\gamma}\right]^{-1}\ .
\een
The choice of an appropriate distance $\Delta R$ has to be done
with care if many multipoles are involved. 
We define $\Delta R$ as the distance to an equidensity surface
$\rho_0(\vec{r})=\rho_{\rm sw}$
of the isovector density $\rho_0$ at some threshold value $\rho_{\rm sw}$.
In practice, we evaluate $\Delta R$ on the grid as
\bea
\label{defDR}
  \Delta R(\vec r_j) &=&
  \quad\min\Big\{\,|\vec r_j-{\vec r_i\!\;}'|\,
  \Big|_{\textstyle\rho_0({\vec r_i\!\;}')\ge\rho_{\rm sw}}\Big\}
  \qquad\mbox{for}\quad \rho_0(\vec r_j)\le\rho_{\rm sw}\nn\\
  \Delta R(\vec r_j) &=&
  -\min\Big\{\,|\vec r_j-{\vec r_i\!\;}'|\,
  \Big|_{\textstyle\rho_0({\vec r_i\!\;}')<\rho_{\rm sw}}\Big\}
  \qquad\mbox{for}\quad \rho_0(\vec r_j)>\rho_{\rm sw}
\eea
where
\ben
  \rho_{\rm sw} = \rho_0^{\rm max}/10\nn
\een
The factor $1/10$ in the definition of
$\rho_{\rm sw}$ has been set by experience. The parameter $\alpha$ is an
effective cut--off distance and  $\gamma$ is the width of the transitional
region; we use here $\alpha=3\,\fm$ and $\gamma=0.4\,\fm$.

Finally, we want to make a few remarks on the numerical procedures
used: We restrict the calculations to axial symmetry. The 
wavefunctions are represented on a grid in cylindrical coordinates
$r=\sqrt{x^2+z^2}$ and $z$. The derivatives are handled as matrices
on the $r$--grid, or $z$--grid respectively. The matrices are
built from a Fourier--definition of the derivatives \cite{deriv}.
We impose no restriction concerning reflection symmetry about
the $z=0$ plane. That is the new feature of the present calculations
compared to earlier work \cite{BluFin,deriv,PRLax}. We are
using a grid spacing of $0.7\,\fm$ in each direction
and are dealing typically with grid sizes of $25\times 75$.
The solution of the field equations is found by interlaced
damped gradient iteration of the nucleon-- and meson--field
equations \cite{Gradit,deriv}. The solution for the Coulomb field
requires a separate handling of the long range parts which reach
far beyond the bounds of the numerical grid; this is done using
the techniques of \cite{FALR}. The iteration includes an
iteration of the constraining force as proposed in \cite{Constit}
and implemented in a relativistic context in \cite{BluFin}.
The iteration scheme has been extended here to deal with two
constraints, which is a straightforward procedure provided
the constraining operators are properly damped outside
the nuclear density, see Eq.~\ref{defDR}.

\section{Fission of heavy elements}
\label{secfiss}

First calculations of fission barriers in the actinides were
published in \cite{Blu94}. They had shown that the relativistic
mean--field model with the parameters NL1 and \plv\ can reproduce
approximately
the double--humped fission barrier of \plut. The first barrier
came out about $4$-$6\,\mev$ higher than the experimental value.
However, it has been shown in the macroscopic--microscopic model
\cite{Nix89} as well as in nonrelativistic Hartree--Fock
calculations \cite{Ber89} that the first barrier is lowered
by about 1-$2\,\mev$ if triaxial deformations are allowed. The
height of the second barrier was about 4-$10\,\mev$ to high in the
previous relativistic calculations. That was due to the restriction
to symmetric deformation. Here we can now present extended
investigations on fission barriers which include also asymmetric shapes
and octupole deformations. We will discuss the nuclei \plut, \radi, and
\thor. And we will compare the PES with those of nonrelativistic
calculations.

\subsection{The fission path for \plut}

In Fig.~\ref{fig50e} we show the PES for asymmetric fission of \plut ,
computed with the quadrupole constraint only.
The octupole deformations have been left free to adjust themselves
to the minimum configuration.
More detailed quantitative information on barriers and minima
is given in Table~\ref{tabpu1}.
The nonrelativistic results with the Gogny force D1s had included an
estimate for correction with the collective zero--point energy, see the
values in brackets in Table~\ref{tabpu1}. These corrections have been
removed for the comparison in Fig.~\ref{fig50e}. The experimental
barriers were obtained by fitting parabolic barriers and minima to the
measured tunneling probabilities. The known part of the zero--point energies
is given in Table~\ref{tabpu1} and corrected for in Fig.~\ref{fig50e}.
Note that these experimental barriers are indirectly determined and
contain an unknown amount of zero--point energies as well as possible
contributions from multidimensional tunneling. Thus a comparison must be
content with a rough proximity of the values. 
The energies and deformations at the minima, the ground state
and first isomeric state, are very well reproduced by the forces
\plv\
and NL1 in comparison to the nonrelativistic results and the data. The
deviations are somewhat large for the force \nls\ but still acceptable.
The height of the first barrier is comparable amongst all theories and
overestimated in relation to the experimental point, even if one
corrects for the triaxial deformation. The second barrier is generally
better reproduced, in particular the force \plv\ comes very close to the
experimental point. In view of the uncertainties on the theoretical as
well as on the experimental side, we see a good agreement between the
relativistic PES from NL1 and \plv\ with the nonrelativistic results and
with the experimental points. The force \nls\ provides also good barrier
heights but the deformation of all minima and barrier is systematically
shifted to lower values compared to all other results. This is probably
due to the higher effective mass of the force \nls. The effect of a
varying effective nucleon mass was studied in \cite{Blu94}: increasing
effective mass softens the  barriers and shifts them to smaller
deformations.

For very large deformations, there is a second branch of solutions
visible in Fig.~\ref{fig50e}. These are strongly favored energetically
at large separations because the fragments are less
deformed internally. These solutions correspond to the fusion valley in the
collective landscape which is distinguished from the fission valley
by a smaller hexadecupole moment \cite{Ber89}. The situation is
analogous for Radium in Fig.~\ref{fig20e}.

Table~\ref{tabpu1} also shows the barriers for \plv\ obtained
from reflection--symmetric calculations \cite{Blu94}. The asymmetric
shapes develop only for larger quadrupole deformations, $Q_2 >
120\,\barn$,
such that only the second barrier is affected. But there the effect
is dramatic and decisive for the fission process which at the end
is dominated by asymmetric fission. We show in Fig.~\ref{fig52e}
the development of the mass asymmetry $A_{\rm h}$ with deformation.
The forces \plv\ and NL1 point strongly towards the main peak in the
experimental mass distribution, denoted "St.~I", whereas the
nonrelativistic calculations with the Gogny force D1s \cite{Ber89}
have their peak at $A_{\rm h}=134$, just at the maximum "St.~II".
It is furthermore interesting to see that the final mass peak is
preformed already at $Q\approx 200\,\barn$ where the two fragments still have
a sizeable overlap. It seems as if the mass flow from right to left and
back is already strongly inhibited although the geometry does not
suggest that. It is most probably a peak in the collective mass
distribution which blocks the flow of matter. This point deserves
further investigations. But these require first a full solution of
the cranking problem along the collective path.

\subsection{The triple--humped fission barrier for \thor}

The PES of \thor\ indicate that there exist strongly stretched
stages because they develop a third fission barrier, see \eg\
the macroscopic--microscopic analysis of \cite{Mol72}. It is
due to strong shell corrections in the outer tail of the
second barrier. The third barrier seems to be experimentally
supported \cite{Blo88}. It is indicated \eg\ by the photofission
cross--section \cite{Zha86}, or by the asymmetric angular distribution
of the light fragment \cite{Bau89}. Fig.~\ref{fig10e} shows
the PES for asymmetric fission of \thor\ for the three relativistic
parametrizations, compared with the barriers from  nonrelativistic
calculations and experimentally deduced barriers.
The details for the barriers and minima are given in
Table~\ref{tab_th}.
It is gratifying to see that all three relativistic parametrizations are
able to reproduce the third minumum and barrier. There seem to be
robust shell effects which appear under widely varying conditions. But
beside this robust pattern, there are now more differences visible
amongst the parametrizations. The force \nls\ behaves somewhat
strangely.
A tendency which was already present in \plut, becomes now even more
disturbing: the first barrier comes out too low and all minima and
barriers are squeezed to lower deformations. The force \nls\ seems to be
not too well adapted for the description of fission PES. Two
explanations are conceivable: First, the problem can come from the
higher effective mass, as it was discussed already in connection with
\plut, and second, the failure at low $Q$ may come from the less carefully
adjusted surface properties. It was observed in connection with the
Skyrme forces that a well--fitted surface tension is required to provide
reasonable fission barriers in the actinides \cite{GueBra}, and we find
in relativistic as well as in nonrelativistic calculations that every
parametrization which fits ground state properties including 
the surface thickness gives comparable and resonable first barriers. The
examples here are the two standard sets, NL1 and \plv, which are
clearly more appropriate. Their overall performance is fair. The second
minimum is a bit to well bound in all cases. But that is a common disease
which is shared with the nonrelativistic models. The two PES of NL1 and
\plv\ develop differences with increasing $Q$. The force NL1 behaves
somewhat better at the second minimum and at the third barrier whereas
\plv\ is superior at the second barrier.  But none of the two is yet
ideal. It is a task for future investigations to search for an even
better force amongst the more versatile stabilized nonlinear
parametrizations of the power--law type (\ref{Ustab}), \ie\ for a better
variant of \plv. Finally, we want to mention that we again see the
fusion valley which is energetically favoured at very large
deformations.

\subsection{Symmetric and asymmetric fission of \radi}

The two previous examples, \plut\ and \thor, prefered asymmetric
fission. However, in the region $84 < Z < 90$ symmetric and
asymmetric fission appear with comparable importance, leading to mass
yields with typically three peaks: the middle peak from symmetric
fission and the two outer peaks corresponding to the light and the
heavy fragment from asymmetric fission. These nuclei should display
an interesting competition between the symmetric and asymmetric PES.
We will consider here \radi\ as a typical example for this region of
nuclei.

The PES for asymmetric as well as symmetric fission of \radi\ are
shown in Fig.~\ref{fig20e}.
The first surprise is that the ground state of \radi\ is asymmetric,
2\,\mev\ lower than the nearby symmetric minimum. But the preference of the
asymetric shape at low $Q$ dissapears quickly. The symmetric shape
has gained already at the first barrier and the first isomeric minimum
is also clearly a symmetric state. Beyond that symmetric and asymmetric
PES develop very differently. The second barrier for asymmetric fission
is much lower but after the subsequent second minumum, a broad
third barrier extends over a wide range of deformations. The second
barrier for symmetric fission, on the other hand, is much higher, but
a second isomeric minimum and an only shallow third barrier follow.
The second symmetric barrier has a further peculiarity: in the range
$120\,\barn < Q <160\,\barn$, there are two branches distinguished by
the hexadecupole moment. It is interesting to note that the 
two branches separate at $Q\approx 120\,\barn$
and merge again at $Q\approx 160\,\barn$. This can be envisaged in the
multidimensional potential energy landscape as having two
bifurcation points at these $Q$ for two distinct minimum pathes embracing
an island of higher energies. The shapes of the fissioning nucleus
for various $Q$ are indicated 
by half--density contours in Fig.~\ref{fig20e}.
The richness of the various shapes deserves a closer look. We show
in Fig.~\ref{fig23e} more detailed plots of the density contours
along the paths.
The features along the symmetric path are particularly remarkable.
Note the strong hexadecupole component of the favoured branch
at the second barrier. And even more impressive are the long and
broad necks for the symmetric configurations at large $Q$.
These will give rise to the observed broad fission mass distributions
due to the possibility of large fluctuations of the actual neck rupture
point \cite{Brosa}.

The problem is that Fig.~\ref{fig20e} does not trivially suggest
a coexistence of symmetric and asymmetric fission. A few further
comments are in order:
\begin{enumerate}
 \item There is only induced fission for \radi\ at rather substantial
  excitation energies, \eg\ from the reaction \radi(p,f) at 11\,\mev\
  \cite{Jen58}; internal excitation reduces the shell effects and thus
  favours symmetric shapes.
 \item The symmetric second barrier will probably be lowered by
  allowing triaxial deformations \cite{Web76}.
 \item It is conceivable that fission proceeds first through
  the asymmetric second barrier and tunnels then towards the
  lower symmetric third minimum at large $Q$.
 \item PES alone can be misleading; tunneling probabilities are
  very sensitive to the collective masses in the tunneling region.
\end{enumerate}
Altogether, we see that the fission of \radi\ is an intriguing problem
which most probably requires a fully fledged collective dynamics
in at least two degrees of freedom, accounting for triaxial shapes,
computing carefully the corresponding collective mass tensor,
and taking care of temperature effects.

Finally, we compare in Fig.~\ref{fig21e} the symmetric PES for the
two relativistic forces NL1 and \plv\ and for a more recent fit
of a Skyrme force \cite{Rei92}.
Table~\ref{tab_ra} complements the detailed information on the
barriers and minima.
We see again the same pattern as in the two previous examples.
The two sets NL1 and \plv\ agree with each other and with comparable
nonrelativistic models for low $Q$ including the first barrier
and perhaps the first isomeric minimum. Differences between NL1 and
\plv\ develop with deformation. The nonrelativistic Skyrme force
lies in between the two relativistic results showing that there
is no principle difference between a relativistic and a nonrelativistic
treatment. The differences of various parametrizations within a class
of models are larger. For example, the force "Skyrme M$^*$", used in the
previous figures, fails to reproduce the third barrier in \radi. An
alternative and more recent Skyrme force from \cite{Rei92} properly
manages to deliver the third barrier. Thus the third barrier may be
used as an additional criterion for selecting effective forces.

\section{Asymmetric ground states in Radium isotopes}
\label{secoctu}

It is an old question whether there exist nuclei which have a ground
state with broken reflection symmetry. Already in the fifties, one
has observed in the actinides low--lying bands of excitations with 
negative parity
\cite{Ste54} which hint at asymmetric deformations in the ground state
\cite{Cha80}. Calculations within the macroscopici--microscopic method
\cite{Mol81,Lea82} as well as nonrelativistic Hartree-Fock calculations
with the Skyrme III force \cite{Bon86} or a Gogny force \cite{rob87} found
asymmetric ground states in the region of Ra--Th which are
by 1-2\,\mev\ lower than the corresponding symmetric ground states.
In this section, we are going to investigate the ground states of
the Ra isotopes.

In Fig.~\ref{fig40e} we show the PES for octupole deformation
of the isotopes of Radium and for the three relativistic
parametrizations discussed in this paper. The symmetric ground state
lies at $Q_3=0$. We see for all three forces that ${}^{216}$Ra
is stable against symmetry--breaking deformation. The softness
against octupole deformation increases with additional neutrons
up to ${}^{226}$Ra, which displays the deepest asymmetric minimum with
the largest octupole moment of all cases. Above ${}^{226}$Ra, the
PES moves slowly back to smaller octupole deformation. Although
the general trends are the same for all three forces, there are
differences in detail, \eg, the transition point changes from $N=220$
for NL1 via $N=222$ for \plv\ up to $N=224$ for \nls.

The PES of Fig.~\ref{fig40e} can be characterized by the octupole
moment at equilibrium and the depth of the minimum compared with the
energy at $Q_3=0$. We compile the information from all isotopes and from
all three forces in
Fig.~\ref{fig41e}.
The figure clearly shows the quantitative difference between the forces.
The force NL1 gives the strongest extra binding, closely followed by
\plv. The force \nls\ has the much softer octupole effects, yielding
even a systematically lower octupole moment for the well deformed
isotopes. The strength of the octupole effects also determines the
transition point from symmetric to asymmetric ground states, the
stronger the effect the earlier the transition. The sequence in the
lower left part of Fig.~\ref{fig41e} reminds one of the sensitivity to the
force which was observed in the transition from spherical to
quadrupole--deformed isotopes of Gadolinium in \cite{Gadpap}. The upper
right
part of the figure also shows the quadrupole moment at equilibrium. It
grows steadily with neutron number and increasing distance 
from the magic number 126. But one can also see that the octupole
deformation has a side effect on the quadrupole deformation because there
is a jump in $Q$ between those forces which have $Q_3=0$ and those with
$Q_3\neq 0$. Finally, in the lower right part of
Fig.~\ref{fig41e}, we show the electrical dipole moment of the ground
state. A nonvanishing dipole moment becomes possible because the
octupole deformations of protons and neutrons are not exactly the same.
The emerging dipole moment seems to be related to the octupole
moment. The force \nls\ with the smaller $Q_3$ also has the
smaller dipole moment. All dipole moments shown here are negative, \ie, 
the protons prefer the bottom of the "pear" whereas the neutrons prefer
the tip. A counterexample are, however, the nonrelativistic results
(see also Tab.~\ref{tab_raz} later on) where the dipole moment has a
different sign for most isotopes. But one has to keep in mind that 
the overall effect is
extremely small. In the upper left part of Fig.~\ref{fig41e}, we have
inserted the experimental energies of the band--heads of the lowest
odd--parity band. These energies should be closely related to the depth
of the octupole minimum. And that is indeed the case, particularly in
comparison with the behaviour of the energies for NL1 or \plv.

We also show in Fig.~\ref{fig41e}
the results from a macroscopic--microscopic model \cite{Sob88,But91,Naz84}.
These
confirm that octupole deformations in the ground state are to be expected
for several Radium isotopes. But the deformations at equilibrium as well
as the transition points differ substantially from the results of the
present relativistic mean--field model. It is to be remarked that these
quantitative details depend strongly on the single particle spectra
near the Fermi energy. It is obvious that the shell model with effective
mass $m^*/m=1$ has a spectrum much different from the mean--field models
all having a rather low effective mass and thus a much lower level
density near the Fermi energy. It is thus interesting to look also
at results from nonrelativistic mean--field models. A comparison
for the two isotopes ${}^{222}$Ra and ${}^{224}$Ra is given in
Tab.~\ref{tab_raz}.
Stable octupole deformations are also found for the Skyrme III force
as well as the Gogny D1s force. But the quantitative details of the
octupole minimum differ. That is not very surprising in view of the fact
that these details depend strongly on the spectral density near the
Fermi energy. And those spectral relations change very sensitively with
the model and even with a slight change of the parameters of one model.
But one should not overinterpret those differences concerning the minima
as we will see from the discussion in the last paragraph of this
section.

As an illustration, we show in Fig.~\ref{fig46e} contour plots of
the ground states of the Radium isotopes, all computed with \plv.
One clearly sees the transition to octupole shapes between ${}^{220}$Ra
and ${}^{222}$Ra as well as the increasing quadrupole deformation.
The figures compare very well with the nonrelativistic mean--field
results \cite{Bon86} and the macroscopic--microscopic method \cite{Sob88}.

In Fig.~\ref{fig48e} we show the relative deviation from the
experimental binding energies with and without asymmetric
degree of freedom for the three relativistic parametrizations
under consideration.
The relative error is always very small, generally below 0.5\% and for
\plv\ even below 0.3\%. Nonetheless it is interesting to see the effect
of the extra binding through octupole deformation at that scale. The
r.m.s.~deviation is slightly improved through asymmetric shapes for NL1
and \plv, but increased for \nls\ which has a bit to much binding
anyway. The figure shows also that there is a significant trend with the
neutron number in the deviations. The fit could not resolve this
isotopic trend in the mismatch. This hints that there is still some open
problem in the model concerning isovector properties. It is to be
noted, however, that we are discussing deviations at a very fine scale
and that the nonrelativistic mean--field models have similar problems
\cite{Rei92}.

Finally, we show in Figs.~\ref{fig30e} and \ref{fig31e} the
PES for ${}^{220}$Ra and ${}^{222}$Ra in the full $Q$--$Q_3$ plane. %
Clearly the octupole deformed minimum in ${}^{222}$Ra is already
announced by the isomeric octupole minimum in ${}^{220}$Ra. The decision
between the minima is related to very small energy changes in a soft
energy surface. This is to be related to the typical $1^-$ excitation
enerergies of 0.5\,\mev, as can be deduced from the upper left part of
Fig.~\ref{fig41e}. That is a typical case where the minimum alone is not
yet conclusive. The true ground state of the system is a coherent
superposition of a large neighbourhood around the minimum, not to forget
the rotational and centre--of--mass projection over the continuum of
energetically equivalent states. The large fluctuations $\Delta Q$ and
$\Delta Q_3$ will add to the ground state deformations for most
observables \cite{ReiDre}, and that can diminuish the differences seen
in Tab.~\ref{tab_raz} or
Fig.~\ref{fig41e}. One needs
first to perform the full collective dynamics in quadrupole and octupole
degrees of freedom before comparing models amongst each other and with
experimental data. That is a far reaching project which has been
accomplished in a few nonrelativistic calculations \cite{Gogcoll,HenFlo}
but which is yet a long way to go for the relativistic models. The
computation of the various PES, as presented in this paper, is a first
step into that direction.

\section{Conclusions}

In this paper we have investigated collective deformation paths for
actinides in the framework of the relativistic mean--field model. The
calculations have been restricted to axial symmetry but allowed for
reflection--asymmetric shapes. The quadrupole moment, damped at the
edges of the numerical box, has been used as constraining force
to generate the fission paths. An additional octupole constraint
had been used to map the collective potential energy surface near
the ground states of Radium isotopes. Three different parametrizations
of the relatistic mean--field model have been employed, NL1 and \nls\
within the standard nonlinear functional for the scalar meson, and
\plv\ as one representative for the stabilized nonlinear functional.

We have studied the collective paths for symmetric and asymmetric
fission of \plut, \thor, and \radi. The relativistic mean--field model
was able to reproduce all the essential qualitative features, as \eg\
the triple--humped barrier in \thor. The two parametrizations NL1 and
\plv\ provide also many quantitative details well in agreement with
comparable nonrelativistic calculations and with experiment. This holds
particularly for the features at lower deformations, \eg\ the ground
state, the first barrier and the first isomeric minimum. It seems that
every mean--field model, which has carefully fitted the nuclear ground
state properties including sufficient information on surface properties,
behaves reasonably in that respect. The force \nls\ falls a little bit
behind, perhaps due to its larger effective mass and/or its lack of
surface information in the fit. Larger differences between the forces
develop for larger deformations. This seems to indicate some need for a further
selection of the forces, or to say it positively, the chance to
discriminate amongst otherwise equivalent forces. In particular the
stabilized nonlinear meson functional has still open degrees of freedom
for a further fine--tuning.

The outer fission barriers are much lower for asymmetric fission than
for symmetric fission in \plut\ and the (asymmetric) fragment masses
from the relativistic calculations agree well with
the data. There is an interesting interplay between symmetric and
asymmetric fission paths in \radi. Although the symmetric barrier
is again much higher, there comes a pronounced third minimum in
the symmetric fission path which may counterweight the barrier
and thus could serve as explanation for the competition between
symmetric and asymmetric fission observed in \radi.

We have furthermore studied the occurence of stable octupole
deformations in the ground states of the Radium isotopes. All three
relativistic forces under consideration have produced such isotopes with
asymmetric ground states. But the transition point from symmetric to
asymmetric shapes is found to depend sensitively on the force. This
holds to some extent also for the octupole moment of the deformed
ground states. A superficial comparison of the ground state energies
with the band head of the low lying odd--parity bands shows that the
forces NL1 and \plv\ provide about the right trends.

In both cases, fission paths and octupole deformations, we have only
looked at the potential energy surfaces from static deformations.
The dynamical aspects, as collective masses and zero--point energy
corrections, have not yet been properly accounted for. This was
appropriate for the present investigations which aimed to explore
first the capability of the relativistic model for the description
of fission and other features of actinides. But now we have reached
an end where further comparisons need to go into more quantitative
detail. This requires as the next important step to implement the
cranking scheme for the computation of the appropriate collective
masses. And honestly, we see presently no simple way to perform
this task in a relativistic environment. Most probably, one has
to perform first a less ambitious step and deduce the collective
masses from the generator--coordinate method.
\section{Acknowledgements}
This work was supported by the Bundesministerium f\"ur Forschung
und Technologie through Contract Number 06OF772 and by the
Joint Institute for Heavy Ion Research.
The Joint Institute for Heavy Ion Research has as member institutions
the University of Tennessee, Vanderbilt University, and the Oak
Ridge National Laboratory; it is supported by the members and by
the Department of Energy through Contract Number DE-AS05-76ERO-4936
with the University of Tennessee.
Two of the authors (P.--G.~R.\ and J.~A.~M.) acknowledge support
by the the Nato Grant CRG.920122.
We wish to express our appreciation to S.~Cwiok and W.~Nazarewicz for fruitful
discussions.

\begin{table}[tbp]
\caption[]{\label{NLsets}
The parameter sets within the standard nonlinear model
used in this work. All masses are
in $\mev$ and $b_2$ is given in $\fm^{-1}$. All other parameters are
dimensionless.}
\begin{tabular}{l*{9}{r@{.}l}}
   & \multicolumn{2}{c}{$g_\sigma$} &
         \multicolumn{2}{c}{$g_\omega$} &
         \multicolumn{2}{c}{$g_\rho$}   &
         \multicolumn{2}{c}{$b_2$}      &
         \multicolumn{2}{c}{$b_3$}      &
         \multicolumn{2}{c}{$m_\sigma$} &
         \multicolumn{2}{c}{$m_\omega$} &
         \multicolumn{2}{c}{$m_\rho$}   & 
         \multicolumn{2}{c}{$m_\nuc$}     \\\hline
NL1    &  10&138 &  13&285 &   4&976 & $-12$&172  & $-36$&265 &
         492&250 & 795&36  & 763&0   &   938&0 \\
\nls\ &  10&444 &  12&945 &   4&383 &  $-6$&9099 & $-15$&8337 &
         526&059 & 783&0   & 763&0   &   939&0 
\end{tabular}
\end{table}
\begin{table}[tbp]
\caption[]{\label{stabsets}
The parameter set \plv\ within the
stabilized nonlinear model. $m_\infty^2$
and $\Delta m^2$ are given in $\fm^{-2}$, $\delta\Phi$ and
$\Phi_0$ are given in $\fm^{-1}$. The nucleon mass and the masses of
the vector mesons are
$m_\nuc=938.9\,\mev$, $m_\omega=780\,\mev$, and $m_\rho=763\,\mev$.} 
\begin{tabular}{l*{7}{r@{.}l}}
   & \multicolumn{2}{c}{$g_\sigma$} &
         \multicolumn{2}{c}{$g_\omega$} &
         \multicolumn{2}{c}{$g_\rho$}   &
         \multicolumn{2}{c}{$m_\infty^2$}&
         \multicolumn{2}{c}{$\Delta m^2$}&
         \multicolumn{2}{c}{$\delta\Phi$}&
         \multicolumn{2}{c}{$\Phi_0$} \\\hline
\plv\ & 10&0514 & 12&8861 & 4&81014 & 4&0 & 3&70015 & 0&269688 &
         $-0$&111914  
\end{tabular}
\end{table}%
\begin{table}
\caption[]{\label{nucmat}
The nuclear matter properties
$E/A \equiv$    binding energy per nucleon,
$\rho_0 \equiv$ equilibrium density,
$K \equiv$      incompressibility,
$m^*/m \equiv$  effective nucleon mass, and
$a_4 \equiv$    symmetry energy
for the relativistic
parametrizations NL1, \plv, and \nls, compared with the two
nonrelativistic Hartree--Fock models, and with the deduced experimental data.
}
\begin{tabular}{l*{5}{r@{.}l}}
  &  \multicolumn{2}{c}{$E/A [\mev]$} & 
     \multicolumn{2}{c}{$\rho_0$ [$\fm^{-3}$]} & 
     \multicolumn{2}{c}{$K [\mev]$} &
     \multicolumn{2}{c}{$m^*/m$} &
     \multicolumn{2}{c}{$a_4 [\mev]$} \\
     \hline
PL-40 & -16&17 & 0&152 & 166&1 & 0&58 & 41&7 \\
NL1   & -16&42 & 0&152 & 211&7 & 0&57 & 43&5 \\
NL-SH & -16&33 & 0&146 & 354&95 & 0&66 & 36&1\\   \hline
Gogny D1s    & -16&32 & 0&166 & 216&0 & 0&67  & 30&8 \\
Skyrme M$^*$ & -16&01 & 0&161   & 219&2 & 0&786 & 30&0 \\  \hline
Exp. &  -15&96 & 0&145 & 240&0  &\multicolumn{2}{c}{} & \empcol 
\end{tabular}
\end{table}%
\begin{table}[tbp]
\caption[]{\label{tabpu1} 
Quadrupole moment of the protons, binding
energy, height of the first ($B_1$) and the second barrier ($B_2$) as
well as of the isomeric state ($M_{II}$) for \plut. 
The parameter sets \plv, NL1,
and \nls\ are compared with nonrelativistic Hartree--Fock calculations
using the forces Gogny D1s \cite{Ber89} and Skyrme M$^*$ \cite{bra85}.
For \plv\ we also show the results of a reflection--symmetric
calculation \cite{Blu94}.
The experimental data are taken from \cite{bem73}, \cite{wap77}, and
\cite{BjoDHB}. The values in parentheses are the zero point energies
subtracted in those references. The second entry in brackets for the
Gogny force D1s gives the lowering of the first barrier due to triaxial
deformations.}
\begin{tabular}{l*{5}{r@{.}l}}
set&
\multicolumn{2}{l}{$Q_{\rm p}[\barn]$}&
\multicolumn{2}{l}{$E_b[\mev]$}&
\multicolumn{2}{l}{$B_1[\mev]$}&
\multicolumn{2}{l}{$M_{II}[\mev]$}&
\multicolumn{2}{l}{$B_2[\mev]$}  \\ \hline
\plv\  &11&9      &$-1812$&1 & 9&5             &0&9       &6&6\\
NL1    &11&7      &$-1811$&9 &10&8            &2&9       &9&4\\
\nls\  &11&1      &$-1818$&8 &8&4             &2&0       &8&5\\ \hline
\plv, sym  &11&8      &$-1812$&0 & 9&7        &1&3       &8&7\\ \hline
Gogny D1s  &\empcol   &\empcol   & 7&$7(+1;+1.7)$    & 2&6  & 7&$5(+1)$ \\
Skyrme M$^*$&\empcol  &\empcol   &11&7            &3&6       &9&1\\ \hline
exp.   &11&$58\pm 0.06$   &$-1813$&5 & 5&$6\pm 0.2$ & 2&$4\pm 0.3$ &
 5&$1\pm 0.2$\\ 
            &\empcol           &\empcol   & \multicolumn{2}{c}{$(+0.5)$} &
\empcol & \multicolumn{2}{c}{$(+0.5)$} 
\end{tabular}
\end{table}%
\begin{table}[tbp]
\caption[]{\label{tab_th} 
Quadrupole moment of the protons,
binding energy, height of the first ($B_1$), the second ($B_2$), and
the third barrier ($B_3$), as well as of the isomeric state ($M_{II}$)
and the third minimum ($M_{III}$) for \thor. All energies are given in $\mev$.
The parameter sets \plv, NL1, and \nls\ are compared with
a nonrelativistic Hartree--Fock calculation using the Gogny force D1s 
\cite{Ber89} and a macroscopic--microscopic calculation
(Yukawa--plus--Exponential and Woods--Saxon) \cite{cwi94}.
The experimental values are from \cite{loeb70}, \cite{wap77}, 
\cite{BjoDHB}, and \cite{Zha86}.
The values in parentheses are the zero--point energies
substracted in those references.}
\begin{tabular}{l*{7}{r@{.}l}}
&
\multicolumn{2}{l}{$Q_{\rm p}[\barn]$}&
\multicolumn{2}{l}{$E_b$}&
\multicolumn{2}{l}{$B_1$}&
\multicolumn{2}{l}{$M_{II}$}&
\multicolumn{2}{l}{$B_2$}&
\multicolumn{2}{l}{$M_{III}$}&
\multicolumn{2}{l}{$B_3$}  \\ \hline
\plv\      &10&0            &$-1763$&6 & 5&9             &$-0$&3 
     &6&7   &1&7   &2&8          \\
NL1         &9&9            &$-1762$&5 &7&1            &1&7    
  &9&5   &5&0   &7&1          \\
\nls\      &8&8            &$-1769$&4 &3&5             &1&4    
  &7&3   &4&4   &6&3             \\ \hline
D1s         &\empcol           &\empcol   & 5&$9(+1)$    & 0&8  
&5&$8(+1)$  &4&2    &4&$3(+1)$       \\
YE+WS       &\empcol           &\empcol   &4&8       &2&1    &6&4  
 &4&2   &8&3          \\ \hline
Exp.$^1$    &9&66   &$-1766$&71 & 5&$8\pm 0.2$ & $\ll 4$&5 & 6&$2\pm 0.2$ 
 &\empcol&\empcol \\ 
            &\multicolumn{2}{r}{$\pm 0.09$}&\empcol
&\multicolumn{2}{r}{$(+0.5)$} &\empcol&\multicolumn{2}{r}{$(+0.5)$}
&\empcol&\empcol\\ 
Exp.$^2$    &\empcol   &\empcol & 5&$82(+0.5)$ & 2&8 & 6&$4(+0.5)$ 
 &\empcol&\empcol \\ 
Exp.$^2(0^-)$ &\empcol &\empcol & 6&$2(+0.5)$ & \empcol & 6&$25(+0.5)$ 
 &\empcol 
&6&$3(+0.5)$ 
\end{tabular}
\end{table}%
\begin{table}[tbp]
\caption[]{\label{tab_ra}
Binding energy, height of the first
($B_1$), the second ($B_2$), and the third barrier ($B_3$), as well as
the height of the isomeric state ($M_{II}$) and the third minimum
($M_{III}$) for \radi. All energies are in $\mev$. 
For each parameter set the upper
line shows the results when only reflection--symmetric shapes are
considered. For \plv, NL1, and LD+MO the barrier heights of the
reflection--asymmetric solution is given in the lower line. "Skyrme" is
a nonrelativistic Hartree--Fock calculation using a recent fit for the
skyrme force \cite{Rei92}, while LD+MO (liquid drop + modified oszillator) 
is a macroscopic--microscopic calculation
\cite{Mol72}.}
\begin{tabular}{l*{6}{r@{.}l}}
&
\multicolumn{2}{l}{$E_b$}&
\multicolumn{2}{l}{$B_1$}&
\multicolumn{2}{l}{$M_{II}$}&
\multicolumn{2}{l}{$B_2$}&
\multicolumn{2}{l}{$M_{III}$}&
\multicolumn{2}{l}{$B_3$}  \\ \hline
    & $-1728$&8 & 6&2     & 1&7     & 16&1 & 2&4 & 5&9 \\
\raisebox{1.5ex}[-1.5ex]{\plv}
    & \empcol   & \empcol & \empcol &  7&6 & 3&7 & 6&3 \\ \hline
    & $-1727$&7 & 6&7     & 2&9     & 19&6 & 8&2 & 11&8 \\
\raisebox{1.5ex}[-1.5ex]{NL1}
    & \empcol   & \empcol & \empcol &  9&6 & 6&4 & 10&5 \\ \hline
Skyrme
    & $-1726$&2 & 7&4     & 3&8     & 18&3 & 7&2 & 9&9 \\ \hline
    & \empcol   & 4&3     & 3&4     & 10&4 & \empcol & \empcol \\
\raisebox{1.5ex}[-1.5ex]{LD+MO}
    & \empcol   & \empcol & \empcol &  9&2 & \empcol & \empcol
\end{tabular}
\end{table}%
\begin{table}[tbp]
\caption[]{\label{tab_raz} 
Properties of the intrinsic ground states
of ${}^{222}$Ra and ${}^{224}$Ra. The results are shown
for the parameter sets \plv, NL1 and \nls\ as well as for 
nonrelativistic calculations with the Skyrme III \cite{Bon86} and the
Gogny D1s \cite{rob87} force and for a macroscopic--microscopic 
calculation (Yukawa--plus--Exponential and Woods--Saxon) \cite{Sob88}.}
\begin{tabular}{l*{2}{cccr@{.}l}}
\multicolumn{1}{c}{} &
\multicolumn{5}{c}{${}^{222}$Ra} &
\multicolumn{5}{c}{${}^{224}$Ra} \\ \hline
& $E_{\rm a}-E_{\rm s}$ & $Q$ & $Q_3$ &
\multicolumn{2}{c}{$D$} &
$E_{\rm a}-E_{\rm s}$ & $Q$ & $Q_3$ &
\multicolumn{2}{c}{$D$} \\
    & $[\mev]$ & $[\barn]$ & $[\barn^{3/2}]$ &
\multicolumn{2}{c}{$[e\cdot\fm]$} &
      $[\mev]$ & $[\barn]$ & $[\barn^{3/2}]$ &
\multicolumn{2}{c|}{$[e\cdot\fm]$} \\ \hline
\plv            & $-0.7$     & 14.8    & 2.4     & $-0$&17 &
                  $-1.4$     & 16.8    & 2.9     & $-0$&4  \\
NL1             & $-1.6$     & 14.9    & 2.5     & $-0$&19 &
                  $-2.1$     & 17.0    & 3.0     & $-0$&47 \\
\nls            &            &         &         & \empcol &
                  $-0.5$     & 15.3    & 2.5     & $-0$&17 \\ \hline
D1s             & $-1.7$     &         & 3.0     &   0&2   &
                  $-1.5$     &         & 3.4     & $-0$&06 \\
SkyrmeIII       & $-0.2$     & 13.0    & 2.0     &   0&04  &
                             &         &         & \empcol \\ \hline
YE+WS           & $-1.0$     & 10.6    & 1.9     & \empcol &
                  $-0.5$     & 11.5    & 2.2     & \empcol 
\end{tabular}
\end{table}%
\cfig{fig50e}{PES of \plut\ for the three parametrizations as indicated.
Barrier heights obtained in
nonrelativistic Hartree--Fock calculations with the Gogny D1s \cite{Ber89}
and the Skyrme M$^*$ force \cite{bra85} as well as experimental
values from \cite{BjoDHB} are drawn for comparison. All barrier
heights are corrected as described in Table~\ref{tabpu1}.}{tbp}
\cfig{fig52e}{Mass of the heavy fragment, $A_{\rm h}$ as defined in
Eq.~\ref{massfrag}, in the fission of \plut\ as function of the
quadrupole deformation $Q_2$. The right hand side shows an experimental
mass distribution which is resolved into two separate mass peaks
\cite{wag89}. The various shapes along the fission path are indicated
by the contours of the vector densities at $\rho_0=0.07\,\fm^{-3}$.
}{tbp}
\cfig{fig10e}{PES of \thor\ calculated with different parameter sets as
indicated. Barrier heights as obtained in a nonrelativistic
Hartree--Fock calculation with the Gogny D1s force \cite{Ber89}, in a
makroscopic--microscopic calculation \cite{cwi94}, and experimental
values are drawn for comparison. exp.$^1$ is taken from \cite{BjoDHB},
exp.$^2$ from \cite{Zha86}.  All barrier heights are corrected for
zero--point energies as described in Table~\ref{tab_th}.}{tbp}
\cfig{fig20e}{Potential energy surfaces of \radi, calculated with the
parameter set \plv. The reference point for the energy is the 
reflection--symmetric ground state. 
Solutions with asymmetric degrees of freedom are
drawn as solid lines and reflection--symmetric solutions are drawn as
dashed lines. The nuclear shapes drawn at their corresponding $Q$ are
lines of constant vector density  at $\rho_0=0.07\,\fm^{-3}$. For the
ground state and the first barrier only the symmetric shapes are
shown.}{tbp}
\cfig{fig23e}{Vector densities $\rho_0$ of \radi\ along the asymmetric
and symmetric fission path, calculated with the parameter set \plv. We
show the same densities here as in \reffig{fig20e}. The contour lines
correspond to the densities 0.03, 0.05, 0.07, 0.09, 0.11, 0.13, 0.15,
and $0.17\,\fm^{-3}$.}{tbp}
\cfig{fig21e}{PES for symmetric fission of \radi\ for the parameter sets
\plv\ and NL1. Additionally the result of a nonrelativistic
Hartree--Fock calculation using a recent fit for the skyrme force 
\cite{Rei92} is
shown.}{tbp}
\cfig{fig40e}{PES for octupole deformation of Radium isotopes for
the three parametrizations \plv, NL1, and \nls. The PES are
symmetric in $Q_3<0$ and $Q_3>0$.}{tbp}
\cfig{fig41e}{Upper left: energy of the octupole minimum relative to the
state with $Q_3=0$ in the isotopes of Radium. In addition
we show the results of macroscopic--microscopic calculations
(Yukawa--plus--Exponential and Woods--Saxon) 
$^1$\cite{Sob88,But91} $^2$\cite{Naz84}. 
The experimental energies of the band--heads of the
lowest odd--parity bands are drawn at positive energies and
indicated by a "1$^-$" \cite{gai83,led78}.
The other subplots show octupole (lower left), quadrupole (upper right),
 and electric dipole
moment (lower right) of Radium nuclei in their 
intrinsic ground state. The deformation parameters $\beta_2$ and 
$\beta_3$ used in \cite{Sob88,But91,Naz84} were converted to 
the quadrupole and 
octupole moments with the approximations 
$\beta_2=\sqrt{5\pi}/(3AR_0^2)\cdot Q$ and
$\beta_3=4\pi/(3AR_0^3)\cdot Q_3$ with
a radius of $R_0=1.2\,A^{1/3}\,\fm$.
The sign of $D$ is shown relative to a positive $Q_3$.}{tbp} 
\cfig{fig46e}{The vector densities $\rho_0$ of Radium nuclei
in their intrinsic ground state, calculated with parameter set
PL--40. Contour lines are drawn for the densities 
0.03, 0.05, 0.07, 0.09,
0.11, 0.13, 0.15 and $0.17\,\fm^{-3}$.}{tbp}
\cfig{fig48e}{Relative difference of the binding energy from
experimental values \cite{wap77} for Radium nuclei. 
On the left hand side the difference is shown for the symmetric
saddle point, on the right hand side for the intrinsic ground
state.}{tbp}
\cfig{fig30e}{PES in the $Q$--$Q_3$ plane for
${}^{220}$Ra calculated with the parameter set \plv.
The distance between solid lines is $0.5\,\mev$ and between
solid and broken lines $0.25\,\mev$.}{tb}%
\cfig{fig31e}{The same as in figure~\ref{fig30e}, but for
${}^{222}$Ra. The intrinsic ground state is denoted
by "$\times$", the symmetric saddle point by "$\bullet$".}{tb}%
\end{document}